\documentclass[aps,pra,twocolumn,groupedaddress]{revtex4-1}  
\usepackage{graphicx}  
\usepackage[caption=false]{subfig}
\usepackage{float}
\usepackage[toc,page]{appendix}
\usepackage{cleveref}
\usepackage{dcolumn}   
\usepackage{bm}        
\usepackage{amssymb}   
\usepackage{color}
\usepackage{amsmath, amsthm, amssymb}
\usepackage{amsfonts}
\usepackage{bbold}       
\usepackage{accents}    
\usepackage{soul}
\newcommand{\beq}{\begin{equation}}
\newcommand{\eeq}{\end{equation}}
\newcommand{\bqa}{\begin{eqnarray}}
\newcommand{\eqa}{\end{eqnarray}}

\newcommand{\erfa}[2]{Eqs.~(\ref{#1}) and (\ref{#2})}
\newcommand{\arf}[1]{Appnedix.~\ref{#1}} 
\newcommand{\crf}[1]{Ref.~\cite{#1}} 

\newcommand{\dg}{^\dagger}

\definecolor{BLACK}{gray}{0}
\definecolor{RED}{rgb}{1,0,0}
\definecolor{GREEN}{rgb}{0.2,.6,0.2}
\definecolor{BLUE}{rgb}{0,0,1}

\newcommand{\blk}{\color{BLACK}}

\newcommand{\sch}{Schr\"odinger}

\newcommand{\sq}[1]{\left[ {#1} \right]}
\newcommand{\cu}[1]{\left\{ {#1} \right\}}
\newcommand{\ro}[1]{\left( {#1} \right)}
\newcommand{\an}[1]{\left\langle{#1}\right\rangle}
\newcommand{\abs}[1]{\left| {#1} \right|}

\newcommand{\ket}[1]{|{#1}\rangle}

\newcommand{\s}[1]{\hat \sigma_{#1}}

\newcommand{\tp}{^{\top}} 

\newcommand{\letter}{paper}

\begin{document}

\widetext


\title{Quantum Jumps Are More Quantum Than Quantum Diffusion}

%

\author{Shakib Daryanoosh}
\author{Howard M.~Wiseman}
\email{H.Wiseman@griffith.edu.au}
 \affiliation{Centre for Quantum Computation and Communication Technology (Australian Research Council),Centre for Quantum Dynamics, Griffith University, Brisbane, Queensland 4111, Australia}

%
%


%
\vskip 0.25cm
\date{\today}

\begin{abstract} 
It was recently argued [\href{http://prl.aps.org/abstract/PRL/v108/i22/e220402}
{Phys. Rev. Lett {\bf 108}, 220402 (2012)}] that 
the stochastic dynamics of an open quantum system  
are not inherent to the system, but rather depend on the existence and nature of a distant detector. 
The proposed experimental tests involved homodyne detection, giving rise to quantum diffusion,  
and required total efficiencies of well over 50\%. Here we prove that for no system is it possible 
to demonstrate detector-dependence using diffusive-type detection of efficiency less than 50\%. 
However, this no-go theorem does not apply to quantum jumps, and we propose a test  
involving a qubit, using different jump-type detectors, with a threshold efficiency 
of only 37\%.
\end{abstract}

\maketitle



Since the advent of quantum trajectory theory some two decades ago \cite{Car93,DalCasMol92,GarParZol92,Bar93,WisMil93c}, 
it has been widely accepted that the stochastic dynamics of individual open quantum systems 
(e.g.~the quantum jumps of atoms) are not inherent to the system, but rather depend upon the presence, 
and nature, of a distant, macroscopic detector (e.g.~a photodiode). However, old ideas, 
namely Bohr's \cite{Boh13} and Einstein's \cite{Ein17} original conceptions of a quantum 
jump as an objective microscopic event, die hard. To disprove these conceptions, or 
indeed any other conceptions of atomic-scale irreversible dynamics as being independent 
of observation (e.g.~that of \crf{GisPer92}) 
it was recently proposed that an experimental test could be performed \cite{WisGam12}. 

The central idea of \crf{WisGam12} was to use two different 
types of detector (e.g.~a photon counter and a homodyne detector) to monitor an 
atom's fluorescence, at a distance, such that the 
two different ensembles of conditioned states could not possibly arise from 
coarse-graining any hypothetical objective pure-state dynamical model 
(OPDM). (It is absolutely crucial to note that the different detectors, 
being distant, do not change the coupling of the atom to the electromagnetic 
field and hence do not change the {\em average} evolution of the atom, 
described by a master equation for its mixed state $\rho$.)
If the efficiency $\eta$ of the two detection schemes were unity then 
they would each produce conditioned states which were pure, and as long 
as some finite measure of those pure states differed between the two ensembles, 
this would prove that there could not be some common pure state dynamics 
underlying them both. Moreover, one can easily recognize that this argument 
is essentially a continuous-in-time version of the Einstein-Podolsky-Rosen (EPR) 
argument \cite{EPR35}. 

Real experiments, however, cannot achieve $\eta=1$, especially 
as one must remember that $\eta$ includes the collection efficiency for the 
atomic fluorescence. Constructing tests to disprove all OPDMs 
for $\eta<1$, as in Ref.~\cite{WisGam12} required  a more general version 
of the EPR phenomenon, introduced in \crf{WisJonDoh07}, and known as 
EPR-steering \crf{CavJonWisRei09,SJWP10} 
(to acknowledge \sch's contribution and terminology \cite{Schro35}).  
In particular, by correlating the continuous measurement record in (say) Alice's distant 
detector,  in some interval $[0,T)$, with the result of various projective measurements performed
directly on the atom by (say) Bob, at time $T$, a carefully constructed EPR-steering inequality
may be tested. (Here $T$ is randomly chosen by Bob.) If the inequality is violated, this proves
that there can be no underlying OPDM for the atom, and thus the stochasticity in its evolution 
(jumps or diffusion) must emanate from the detectors. 

In \crf{WisGam12}, the system considered was a strongly driven two-level atom, 
and an appropriate EPR-steering inequality was chosen. The best violation was found using the following pair of 
 monitoring schemes, also known as {\em unravellings}: 
i) homodyne detection, giving rise to diffusion; and ii) photon counting, using spectral filtering and a weak local oscillator (LO) with 
adaptively controlled phase, giving rise to jumps. Assuming equal efficiencies for both detectors, the critical efficiency 
required was $\eta_c \approx 0.58$. Moreover, replacing the (extremely complicated) scheme (ii) with 
a different homodyne scheme raised this to $\eta_c \approx 0.73$.

In this \letter, we show that the high efficiency required for tests involving homodyne detection is no 
accident: for any system, no matter the number $L$ of outputs, and no matter the number $M$ of 
different unravellings, if they are all diffusive and all efficiencies are below $0.5$, 
it is impossible to demonstrate EPR-steering. Following the proof of this no-go theorem, 
 we show that it does not apply to jump unravellings by exhibiting a qubit system, 
with $L=2$ and $M$ large, in which $\eta_c \approx 0.37$. 
Moreover, even restricting to $M=5$, a decent (5\%) 
violation is predicted for efficiency $\eta_d \approx 0.455$, whereas under the same 
conditions with diffusive unravellings 
the best we could obtain was $\eta_d \approx 0.78$. That is, the peculiarly quantum nature 
of open systems are far more easily manifest by quantum jumps than by quantum diffusion.


\section{Open quantum systems and diffusive unravellings} 
We restrict to Markovian systems, since non-Markovian quantum systems do not, in general, allow 
for pure conditioned states even for 100\% efficient non-disturbing detection \cite{WisGam08}. 
Then the average, or unconditioned, evolution is described by a master equation (ME) 
\beq \label{GME}
\dot{\rho} =  -i[\hat{H},\rho] + 
{\cal D}[\hat {\bf c}] \rho 
\equiv {\cal L}\rho.
\eeq
Here $\hat {\bf c} = (c_1,\cdots,c_L)^\top$  is an arbitrary vector of operators (called Lindblad  operators), 
and ${\cal D}[\hat {\bf c}]\equiv \sum_{l=1}^{L} {\cal D}[{\hat{c}}_l] $,
where ${\cal D}[\hat{c}]\rho\equiv \hat{c}\rho{\hat{c}}\dg-1/2({\hat{c}}\dg\hat{c}\rho+\rho{\hat{c}}\dg\hat{c})$  \cite{WisMil10}. 
This equation results from tracing over that environment to which the system is coupled, 
but it is possible to monitor the environment and get further information about the system. 
This results in a conditioned state $\varrho$ which is (in general) more pure, and which evolves 
stochastically according to the measurement record. Different ways of monitoring the environment 
give rise to different unravellings of the ME. For example, 
in quantum optics, a LO of arbitrary phase and amplitude may be added 
to the system's output signal prior to detection. For a 
weak LO (i.e.~one comparable to the system's output field), individual photons may be counted, 
giving rise to quantum jumps in $\varrho$, but for a strong LO only a photocurrent is recorded, 
giving rise to quantum diffusion in $\varrho$ \cite{Car93,WisMil10}.




 The most general diffusive unravelling of Eq.~(1) is described by the 
stochastic ME \cite{WisMil10}
\beq \label{Gdif}
d\varrho = dt {\cal L} \varrho + {\cal H}[d{\bf Z}\dg(t) \hat {\bf c}] \varrho,
\eeq 
where  
${\cal H}[\hat{a}]\rho\equiv \hat{a}\rho+\rho\hat{a}\dg-{\text Tr}[\hat{a}\rho+\rho\hat{a}\dg]\rho$, 
and $d{\bf Z}(t)$ is a vector of c-number Wiener processes. Physically, these arise as noise 
in photocurrents, and have the correlations
$d{\bf Z} d{\bf Z}\dg = {\Theta} dt$ and  $d{\bf Z} d{\bf Z}^\top = \Upsilon dt$.
Here ${\Theta} = {\rm diag}(\eta_1,\cdots,\eta_L)$ is a 
real diagonal matrix, with $0\leq \eta_l\leq 1$ being the efficiency with which output channel $l$ 
is monitored.
The complex symmetric matrix $\Upsilon$, on the other hand, parametrizes all of the 
diffusive unravellings this allows. It is a subject only to the constraint that the {\em unravelling matrix} \blk 
\beq  \label{formU}
U(\Theta,\Upsilon) \equiv  \frac{1}{2}\left( 
\begin{array}{cc}
\Theta+\text{Re}\left[ \Upsilon \right]  & \text{Im}\left[ \Upsilon\right]  \\ 
\text{Im}\left[ \Upsilon\right]  & \Theta-\text{Re}\left[ \Upsilon\right] 
\end{array}
\right),
\eeq
 be Positive Semi-Definite (PSD), i.e.~$U(\Theta,\Upsilon) \geq 0.$

Consider two different unravellings 
$ U$ and $ U^0$. If it is possible to write
$d{\bf Z}^0 = d{\bf Z} + d\tilde{\bf Z}$, where $d\tilde{\bf Z}$ is an 
unnormalized complex vector Wiener process uncorrelated with $d{\bf Z}$, then clearly 
the first unravelling $U$ can be realized by implementing the second, 
$U^0$, and throwing out some of the information in record. 
This will be the case if the implied unravelling matrix for $d\tilde{\bf Z}$, 
$\tilde{U} \equiv U^0-U$, is PSD, in which case we call  
$U$ a {\em coarse-graining} of $U^0$. 

Now say that Alice can implement a set 
$\{U^m\}_{m=1}^{M}$ of different unravellings of the form of 
Eq.~(3).  A {\em necessary} condition for this set to be capable of demonstrating continuous 
EPR-steering,  is that they not be coarse-grainings of a single 
unraveling \cite{WisGam12}; that is, that there not exist a $U^0$ such that 
$\forall m, U^0-U^m \geq 0$, since the stochastic evolution defined by 
$U^0$ would be an explicit OPDM compatible with all of the observed behaviour. 
From this, we can show the following: \\
{\bf Theorem} [No-go for inefficient diffusion]. 
If, for a set of diffusive unravellings $\{U^m\}_{m=1}^{M}$ of an 
arbitrary ME (1), 
the efficiencies satisfy $$\forall m, \forall l, \eta_l^m \leq 0.5,$$ 
then this set cannot be used to demonstrate the detector-dependence 
of the conditional stochastic evolution. 

 \begin{proof} Consider $U^0 = U(I,0)$ [as per Eq.~(3)]. Then under the condition of the theorem, 
 $\tilde U^m \equiv U^0-U^m$ equals $U(I-\Theta^m, -\Upsilon^m) = U(\Theta^m,-\Upsilon^m) + \Xi^m$, 
 where  $\Xi^m={\rm diag} (\varepsilon_1^m,\cdots,\varepsilon_L^m,\varepsilon_1^m,\cdots,\varepsilon_L^m)$,
 where $\varepsilon^m_l \equiv 1 - 2\eta^m_l$ is non-negative for all $l$ and $m$. 
 To establish the result we need prove only that $\lambda_{\rm min}(\tilde{U}^m) \geq 0$ 
 for all $m$.  Using Weyl's inequality \cite{Ber05}, 
 $\lambda_{\rm min}\big[\tilde{U}^{m}] \geq \lambda_{\rm min}\big[U(\Theta^m,-\Upsilon^m)\big]+{\rm min}_l\cu{\varepsilon_l^m}$.
  It can be also proven, based on the properties 
of partitioned matrices \cite{Ber05}, that $\cu{\lambda[U(\Theta^m,-\Upsilon^m)]} = \cu{\lambda[U(\Theta^m,\Upsilon^m)]}$. 
Since $U(\Theta^m,\Upsilon^m) \geq 0$ by definition, the result follows. \end{proof}

To reiterate: unless it is the case that for at least one output channel, and at least 
one unravelling, the monitoring efficiency is greater than $0.5$, then there exists an unravelling $U_0=U(I,0)$, 
which defines an OPDM which is consistent with all the observed conditional behaviour of 
the system, so that no detector-dependence can be proven. (It is interesting to note that this model, corresponding to 
the unravelling $U(I,0)$ is precisely that introduced, without a measurement interpretation, 
as quantum state diffusion in Ref.~\cite{GisPer92}.)  This is the first main result of our \letter. 
The second is that this condition, $ \eta_l^m \blk > 0.5$, is {\em not} necessary for 
quantum jump unravellings, as we now show. \blk

\section{Evolution via quantum jumps} A more general class of unravellings 
(in that it contains quantum diffusion as a limiting case \cite{WisMil10}) is that of quantum jumps, 
whereby the conditioned evolution of the system undergoes a discontinuous change upon 
certain events (``detector clicks''), 
and otherwise evolves smoothly \cite{WisMil10}. 
There is not just one jump unravelling; 
for the general ME (1), for instance, each output channel can have a weak 
 LO added to it prior to detection \cite{WisMil10}. 
When a click is recorded in the $l$th output in interval $[t,t+dt)$, the system state is updated via 
\beq
\varrho(t) \to \tilde{\varrho}_l(t+dt)  = dt\eta_l{\cal J}[\hat c'_l]\varrho(t).
\eeq
Here 
${\cal J}[\hat{a}]\varrho \equiv \hat{a}\varrho \hat{a}\dg$, and 
the jump operator is $\hat c'_l = \hat c_l + \mu_l$, where $\mu_l$ is an arbitrary complex number. 
The norm of the unnormalized state $\tilde{\varrho}_l$ is equal to the probability of this click. 
If no click is recorded the system evolves via \cite{WisMil10}
\beq
\varrho(t) \to \tilde{\varrho}_0(t+dt) = \sq{1+dt{\cal L} - dt\sum_{l=1}^{L} {\eta_l} {\cal J}[\hat c'_l]}\varrho(t),
\eeq
as required for the Eq.~(1) to be obeyed on average, where again the norm is equal to the 
probability 
of their being no clicks in that infinitesimal interval \cite{WisMil93c}.
  
 In the case of efficient detection, the panoply of jumpy unravellings bestows an 
 extraordinary power upon the experimenter: to confine the conditioned state of the system, 
 in the long-time limit of an ergodic ME, to occupying only finitely many 
 different states in Hilbert space \cite{KarWis11}. Such a set of states, with the probabilities 
 with which they are occupied, as in $\cu{(\wp_k,\ket{\phi_k})}_{k=1}^K$,  is called 
 a physically realizable ensemble (PRE) \cite{WisVac01}. For the case of a qubit, a PRE of 
 the minimum size ($K=2$) always exists \cite{KarWis11}. Moreover,  there are as many $K=2$ PREs 
 as there are distinct (not necessarily linearly independent) real eigenvectors of the matrix $A$, 
 which appears in the Bloch equation 
$\dot{\vec{r}}=A\vec{r}+\vec{b}$ equivalent to Eq.~(1) for the qubit state represented 
by $\vec{r} = \an{(\s{x},\s{y},\s{z})\tp}$
 
\begin{figure}
\captionsetup[subfigure]{labelformat=empty}
\subfloat[]{\includegraphics[scale=0.7]{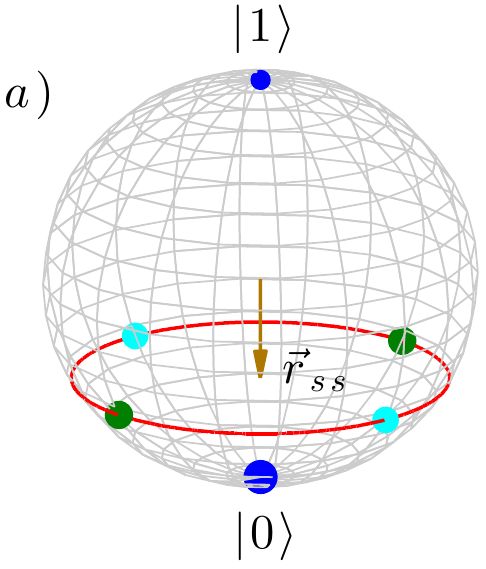} \label{fig:Bloch} } \,
\subfloat[]{\includegraphics[scale=0.495]{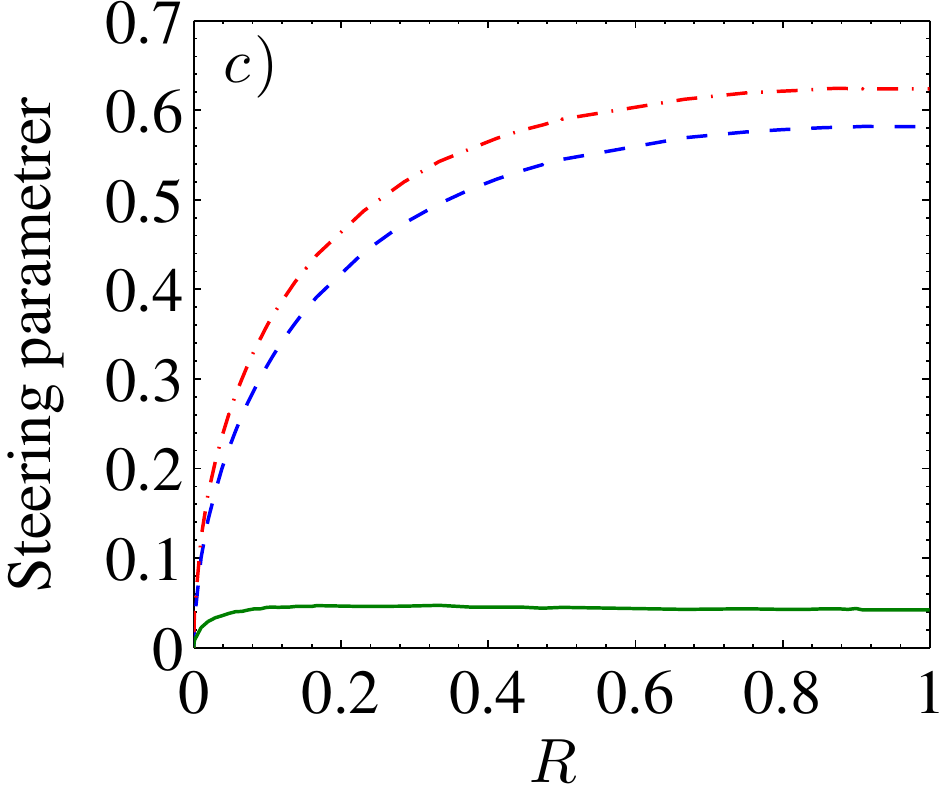} \label{fig:VSetaj}} \\[0.01pt]
\subfloat[]{\includegraphics[scale=0.69]{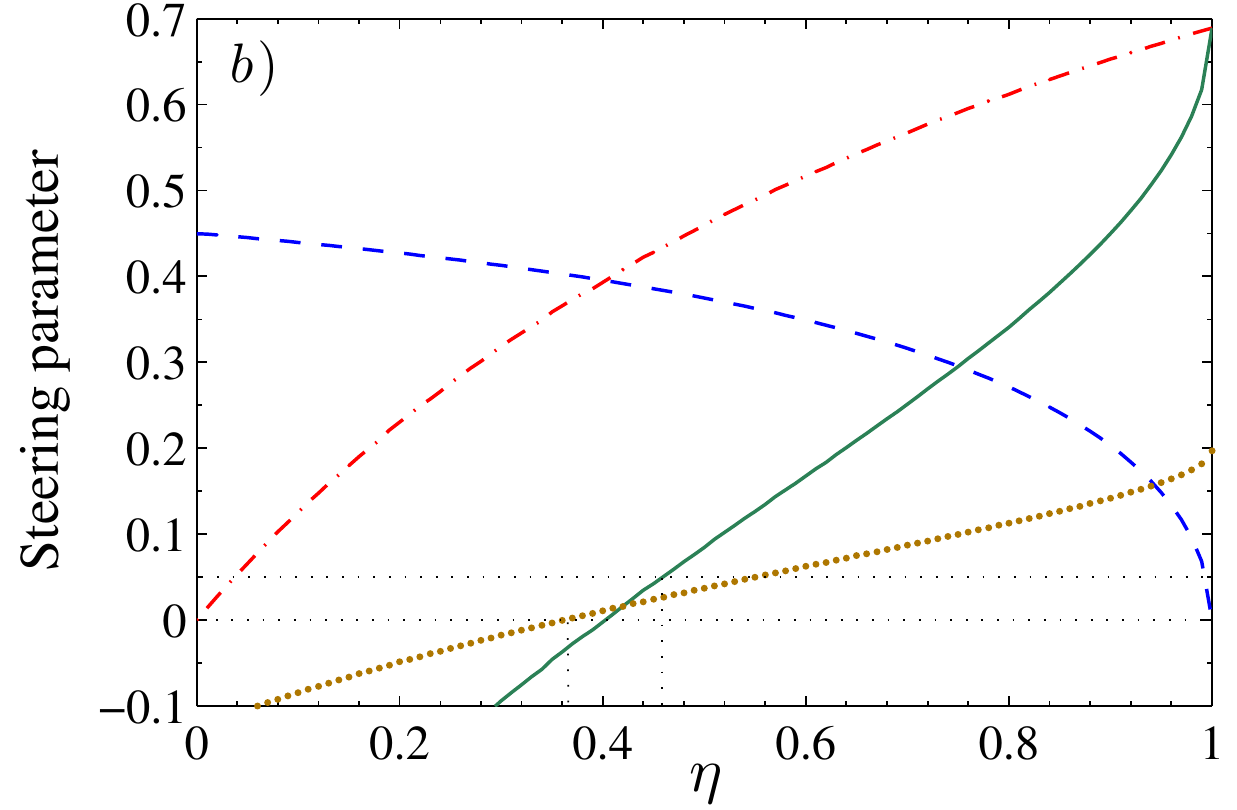} \label{fig:vsRj} } 
\caption{\label{fig:jump} (Color online). Quantum jump unravellings: (a) Bloch sphere representation for the case $R=\gamma_+/\gamma_- = 1/3$ of the steady state $\vec{r}_{ss}$ (brown arrow), and three different $K=2$ PREs: $E^z$ (dark blue) at the poles, and on the equator $E^{\varphi=0}$ (dark green), and $E^{\varphi = \pi/2}$ (light blue). In all cases the volume of the spheres represent the probability of that pure state in the PRE being occupied. 
The red dashed circle show the locus of all the $E^\varphi$s. 
(b) The value of the steering parameter (8), versus $\eta$: for $n\to \infty$ 
settings at $R=0.01$ (brown dotted); and for $n=4$ and $R_o= 0.16$ (green solid). 
For the latter case 
the first term of the EPR-steering inequality (8) (red dot-dashed), and 
the second term of (8) (blue dashed), are also shown. (c) displays the steering parameter $S$ and the same ensemble averages as in (b), with the same line styles,
as a function of $R$ for efficiency $\eta_d= 45.5\%$. 
 $S$ 
 has maximum value of $0.05$ at $R_o= 0.16$. } 
\end{figure}

Consider now a particular qubit ME with $L=2$ \blk irreversible channels into the environment, 
as follows: 
\beq \label{ME}
\dot{\rho} = \gamma_- {\cal D}[{\hat{\sigma}}_-] \rho + \gamma_+ {\cal D}[{\hat{\sigma}}_+] \rho,
\eeq 
where $\s{\pm} = (\s{x}\pm i\s{y})/2$  are raising and lowering operators respectively. That is, 
in the notation of Eq.~(1), $\hat H=0$ (in a suitable rotating frame), and $\hat c_\pm = \sqrt{\gamma_\pm}\s{\pm}$. 
Note this ME is completely different from the single decoherence channel of \crf{WisGam12}. 
 Realization of this sort of ME has been recently investigated in the context of quantum computing, in the limit 
 $\gamma_+=\gamma_-$, for which suitable unravellings allow universal computation to be performed \cite{SanCar12}.
The Bloch representation of Eq.~(6) is $\dot{\vec{r}}=A\vec{r}+\vec{b}$ with 
\beq  \label{BlochA}
A = -\gamma_\Sigma \left( 
\begin{array}{ccc}
1/2  & 0 & 0  \\ 
0 & 1/2 & 0 \\
0 & 0 & 1 
\end{array}
\right)\;,\;\; 
\vec{b}=\gamma_\Delta \ro{\begin{array}{c} 0 \\ 0 \\ 1 \end{array}}\;,
\eeq
where $\gamma_\Sigma=\gamma_+ + \gamma_-$ and $\gamma_\Delta = \gamma_+ - \gamma_-$. 
This $A$ has one real eigenvector in the $z$ direction, giving rise to a \blk $K=2$ PRE, 
expressed in the Bloch representation $\{(\wp_{\pm}, \vec{r}_{\pm})\}$ as 
$E^z\equiv\{ (\gamma_\pm/\gamma_\Sigma, (0,0,\pm1)\tp) \}$. 
This $A$ also has infinitely many real eigenvectors in the $x$--$y$ plane, which we can parametrize 
by the azimuthal angle $\varphi$, giving rise to the $K=2$ PREs 
$E^\varphi \equiv\{(1/2,  (\pm C \cos\varphi, \pm C \sin\varphi, z_{\rm ss})\tp)\}$
where $C=2\sqrt{\gamma_+\gamma_-}/\gamma_\Sigma$ and  $z_{\rm ss}= 
\gamma_\Delta/\gamma_\Sigma$. All of these ensembles average to give the steady-state 
Bloch vector $\vec{r}_{\rm ss}=(0, 0, z_{\rm ss})$ as shown in Fig.~\ref{fig:Bloch}.

Following Ref.~\cite{KarWis11} we can determine the LO amplitudes 
to realize these PREs (assuming perfect detection). For $E^z$ it is trivial to see that no LO is required, as 
$\hat c_\pm$ cause jumps between the $\s{z}$-eigenstates. For $E^\varphi$ we require 
an {\em adaptive} scheme with each $\mu^\varphi_l(t)$ taking two possible values, 
$\mu_l^\varphi{}^\pm = \pm   \sqrt{\gamma_{-l}} \, e^{l i \varphi}/2$.  
Here $l$, the label for the output channel, also takes the value $\pm$, but that is independent 
of the $\pm$ defining the two values for the LO. The adaptivity required is that every time 
a detection in {\em either} channel occurs, the LO for {\em both} channels is swapped from their $+$ values 
to the $-$ values, or {\em vice versa}.  

\section{Quantum jumps are more loss-tolerant} 
For the unit-efficiency case, the $z$ unravelling (i.e.~that giving rise to ensemble $E^z$) 
is such that $\an{\s{z}}^2=1$. 
If this unravelling were the OPDM of the system then 
the complementary variable $\s{\varphi}=\s{-}e^{i\varphi}+\s{+}e^{-i\varphi}$ (for any $\varphi$) would 
necessarily have zero mean, but we know that it has a nonzero conditional 
mean for the PRE $E^\varphi$. 
Consider a finite set of $n$ different $\varphi$ values $\cu{\varphi_j=(j/n)\pi}$, 
so that, with the $z$ unravelling, Alice has a total of $M=n+1$ unravellings. 
Then the above, unit-efficiency, considerations suggest the following EPR-steering 
inequality \cite{JonWis11}:	 
\beq  \label{eqn:eprs}
 S\equiv \frac{1}{n}\sum_{j=1}^n {\rm E}^{\varphi_j} \Big[\abs{\an{\hat{\sigma}_{\varphi_j}}}\Big]
  -  f(n){\rm E}^z\sq{ \sqrt{1-{\langle {\hat{\sigma}}_{z} \rangle}^2}} \le 0 .
\eeq 
Here ${\rm E}^z[\bullet]$ means the ensemble average under the $z$ unravelling, 
so that $\langle {\hat{\sigma}}_{z} \rangle$ appearing therein means ${\rm Tr}[\varrho{\hat{\sigma}}_{z}]$, 
where $\varrho$ is  the conditional state $\varrho$ under that unravelling, and likewise for the $\varphi_j$ unravellings.
The function $f(n)$ is defined in Ref.~\cite{JonWis11} and asymptotes
to $2/\pi$ as $n\to \infty$. 

In Eq.~(8) we are not assuming 
unit efficiency; the unravellings are as defined above, but the long-time 
conditional states will not be pure (and will certainly not be just two in number for each unravelling).  
Our aim is to show that the no-go theorem for inefficient detection, 
applicable to diffusive unravellings, is not universal, by showing that Eq.~(8)
can be violated for $\eta < 0.5$. To do this we must evaluate the terms on the LHS 
for the $n+1$ different unravellings, although we note that by the symmetry of the problem, 
${\rm E}^{\varphi} [|\an{\hat{\sigma}_{\varphi}}|]$ is independent of $\varphi$. 
The ensemble average ${\rm E}^z$ can be done semi-analytically, while that for 
${\rm E}^\varphi$ requires stochastic simulation; see Appendix \ref{appnA}. We plot these averages in 
Fig.~\ref{fig:jump}, as well as $S$ in Eq.~(8), for varying $R\equiv \gamma_+/\gamma_-$, 
and varying $\eta$, and for $n=4$ and $n=\infty$. 


The critical threshold efficiency for jumps to violate Eq.~(8), 
 is $\eta_c \approx 0.37$, which is considerably below the limit of $0.5$ necessary for diffusion 
 according to our Theorem. This is the second main result of this \letter. This $\eta_c$ is 
 achieved in the limits $R\ll 1$ and $n \to \infty$, neither of which are convenient because 
 the first implies that even when there is a violation it will always be very small (compared to the maximum 
 possible violation of unity at $\eta=R=1$), and the second 
 because it requires infinitely many measurement settings. However, we show that a decent violation, 
 of 0.05, is achievable with only $n=4$, with an efficiency $\eta_d \approx 0.455$,  which is still 
 significantly below the $50\%$ limit. This was for an optimized value of $R$, found numerically, 
 of $R_o\approx 0.16$.

\section{Comparison with quantum diffusion} 
\begin{figure}
\includegraphics[scale=0.69]{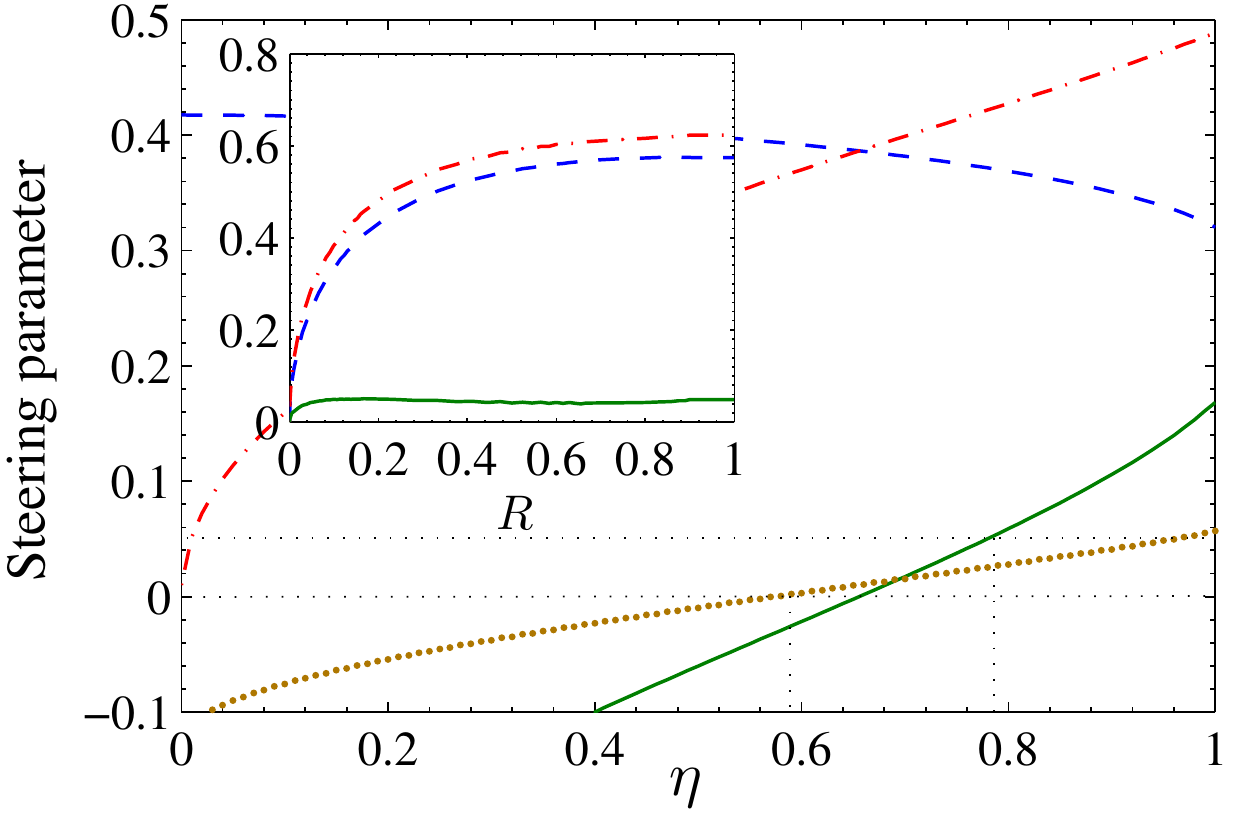}
\caption{\label{fig:diff} (Color online). Quantum diffusion unravellings: The value of the steering parameter (8), versus $\eta$: for $n\to \infty$ 
settings at $R=0.01$ (brown dotted); and for $n=4$ and $R_o=0.13$ (green solid).
Also shown are 
the first and second terms of the EPR-steering inequality (8), both for $n=4$ (red dot-dashed and blue dashed, respectively).
Inset displays the steering parameter $S$ 
and the same ensemble averages as in the main figure, with the same line styles,
as a function of $R$ for efficiency $\eta_d=78\%$. 
 $S$ has maximum value of $0.05$ at $R_o=0.13$. }
\end{figure}

We now consider the same ME (6) and the same 
EPR-steering inequality (8), applied to diffusive unravellings (2). 
Here, with $\hat{\bf c} = (\sqrt{\gamma_-} \hat{\sigma}_-, \sqrt{\gamma_+} \hat{\sigma}_+)^{\top}$ 
we have  $\Theta = {\rm diag}(\eta,\eta)$ and the optimal $\varphi_j$ unravelling is $\Upsilon = \eta \, {\rm diag}(e^{-2i\varphi}, e^{2i\varphi})$. \blk 
For diffusive unravellings, there is no unravelling that is particularly useful for Alice to be able to predict Bob's 
value for $\hat\sigma_z$, so for the $z$ unravelling we simply use an arbitrary $\varphi$ unravelling (this is still better
than using no unravelling i.e.~replacing the ${\rm E}^z[\bullet]$ term by $\sqrt{1-z_{\rm ss}^2}$). Since, 
as in the jump case, ${\rm E}^{\varphi} [|\an{\hat{\sigma}_{\varphi}}|]$ is independent of $\varphi$, 
we only have to simulate one unravelling. This is described in \arf{appnA} and the results are shown in Fig.~\ref{fig:diff}. 

The critical efficiency is $\eta_c \approx 0.59$, greater than $0.5$ as expected, for $R \ll 1$. 
While this is less than the all-diffusive $\eta_c =0.73$ of \crf{WisGam12}, it is a long way above  
the quantum jump $\eta_c = 0.37$ found above. Interestingly, analytical calculations (see \arf{appnA}) 
show that for $R\ll 1$, $S \cong g(\eta) \sqrt{R}$, where $g(\eta)$ changes sign at $\eta_c$. 
Restricting to $n=4$ and looking again for a decent ($0.05$) violation, 
we obtain $\eta_d = 78\%$ at an optimal value of $R_o=0.13$.

In conclusion, for the experimental task of ruling out all objective pure-state dynamical models for an 
open quantum system 
we have: (i) proven it is impossible to achieve this by diffusive unravellings with efficiencies below 50\%; and 
(ii) exhibited a set of quantum jump unravellings that would allow such a task, for a qubit, 
with an efficiency as low as $\eta_c =  37\%$. Moreover, even allowing for a decent margin of error and other 
experimental realities, a jump efficiency of only $\eta_d = 45.5\%$ is required for our system, 
whereas the corresponding figure for 
diffusive unravellings is $\eta_d = 78\%$. That is, it is far easier to show that the stochasticity of 
quantum jumps arises in the distant detector (as opposed to being intrinsic to the system)  
than it is to show this for quantum diffusion, and in that sense the former are more quantum. 
For future work we believe that it will be possibly to prove even stronger no-go theorems 
for diffusive unravellings. But we also hope that better EPR-steering tests may allow 
experimentalists to get closer to the limits of such no-go theorems than our results here, 
so that the recently reported diffusive monitoring efficiency of 
$49\%$ in superconducting qubit experiments \cite{Mur13} is encouraging.

This research was supported by the ARC Centre of Excellence Grant No. CE110001027. We thank Jay Gambetta for discussions.   

\blk

\appendix
\numberwithin{equation}{section}
\section{Supplemental Material} \label{appnA} 
\subsection*{ Simulating Averages for Adaptive Detection}
Let's say system starts at time $t=0$ from some arbitrary initial state. 
Based on the quantum trajectory theory relevant for inefficient detections $\eta<1$, the unnormalized system state matrix $\tilde{\varrho}$ then evolves according to \cite{WisMil10a}
\beq \label{nojump}
\dot{\tilde\varrho}(t) =  {\cal  L}  \tilde \varrho - \eta \big( {\cal J}[\hat{c}_-^{\prime}] + {\cal J}[\hat{c}_+^{\prime}] \big) \tilde\varrho,
\eeq
where $\hat{c}_{\pm}^{\prime} = \hat{c}_{\pm} + \mu_{\pm}$ with $\mu_{\pm}$ the  
weak local oscillator amplitudes, and 
${\cal J}[\hat{a}]\bullet =\hat{a}\bullet \hat{a}\dg$. Until the first jump occurs at time $t=t_1$ the 
system's state is described by the conditional state $\varrho(t)=\tilde{\varrho}(t)/\text{Tr}[\tilde{\varrho}(t)],$ where $\text{Tr}[\tilde{\varrho}(t)]$ is the probability of no photon detection for the interval $[0, t_1)$. 
We can generate the jump time $t_1$ with the correct statistics by generating a random number $u$ uniformly distributed in $[0,1]$, and solving $\text{Tr}[\tilde{\varrho}(t_1)] = u$. Then using  a new random number 
we determine which jump occurs with the relative weights $w_\pm = \eta{\rm Tr}[\hat{c}_{\pm}^{\prime}{}
\dg\hat{c}_{\pm}^{\prime}\varrho(t_1)]$. The new starting state of the system is  
$\varrho'(t_1) = \eta{\cal J}[\hat{c}_{\pm}^{\prime}]\varrho(t_1)/w_\pm$ 
corresponding to the appropriate jump operator. 
This algorithm is repeated to generate a long sequence of subsequent jumps at times $t_2, t_3, \cdots$,  
and once the transients have decayed away 
we can   
obtain the required ensemble averages as time averages:  
\beq \label{avbeta}
{\rm E}^\varphi \Big[|{ \langle {\hat{\sigma}}_{\varphi} \rangle}| \Big] = \lim_{N \rightarrow \infty} \frac{1}{t_N-t_n} \sum_{j=n}^{N}  \int_{t_{j-1}}^{t_j} \blk |{\text{Tr}[ \hat{\sigma}_{\varphi} \varrho(t)]}| \, dt .
\eeq \blk
Here $n\sim 10$ is the number of jumps after which system can be taken to have relaxed to the steady state, and  $t_j$ is the time of each individual jump. We use $N-n= 10^4$.

\subsection*{ Calculating Averages for Direct Detection}
Under direct detection (zero local oscillator), each jump operator $\hat c_\pm$ prepares the system in one of the states $\vec{r}_{\pm}=(0,0,\pm1)$, and this 
fact enables us to solve for this dynamics without resort to stochastic simulation. Following a jump, 
the unnormalized system state again smoothly evolves, as in Eq.~(A.1) but this time without any 
 local oscillator field, that is
\beq \label{nojumpz}
\dot{\tilde\varrho}(t) =  {\cal L} \tilde \varrho - \eta \big( {\cal J}[\hat{c}_-] + {\cal J}[\hat{c}_+] \big) \tilde\varrho.
\eeq
Here the conditional state matrix $\varrho(t)=\tilde{\varrho}(t)/\text{Tr}[\tilde{\varrho}(t)]$ until the first jump is a mixture of $\vec{r}_+$ and $\vec{r}_-$. Depending on the initial condition, these states are 
$\tilde{\varrho}^{\pm}(t) = \frac{1}{2} [p^\pm(t)+\tilde{z}^\pm(t)\s{z}]$, 
where $p^{\pm}$ and $\tilde{z}^{\pm}$ are the solution of the following set of equations   
{\setlength\arraycolsep{2pt}
\begin{eqnarray}  {\label {dd}}
\dot{p} & = & -\eta/2 (\gamma_{\Sigma} p + \gamma_{\Delta} \tilde{z}) , \\[4pt] 
\dot{\tilde{z}} & = & (\eta/2 - 1) [\gamma_{\Sigma} \tilde{z} + \gamma_{\Delta} p).
\end{eqnarray}
with the appropriate initial condition of $p^\pm(0) = 1$ and $\tilde{z}^\pm(0)=\pm1$.   \blk

Then the system jumps into either of $\vec{r}_{j}$ with rates \blk $w^\pm_{j}(t)=\eta{\rm Tr}[{\hat{c}_{j}\dg} \hat{c}_{j}\varrho^\pm(t)]= \eta [1+j\times z^\pm(t)]/2$, where $j=\pm$ also. 
This is such that the probability that, when a jump occurs, the system jumps into state $\vec{r}_\pm$ 
is  $\wp_\pm$ and can be obtained by solving
\beq \label{wpell}
\wp_j = \sum_{\ell=\pm} \wp_\ell \int_0^\infty p^\ell(t) w^\ell_j(t)\, dt.
\eeq
 Here $\wp_\ell$ appears as the probability for starting in state $\vec{r}_\ell$ at some time $t_{\rm jump}$, and 
$p^\ell(t) w^\ell_j(t)\, dt$ is the probability that, given this starting point, a jump occurs in the  interval   
$[t_{\rm jump}+t, t_{\rm jump}+t+dt)$ and puts the system into state $\vec{r}_j$. Averaging over the two possible initial states and all the possible times 
from one jump to the next, should give $\wp_j$ (i.e. the same function as $\wp_\ell$, since nothing distinguishes
the first jump from the second in the long-time limit.) 

Solving Eq.~(A.6) analytically gives $\wp_{+} = \wp_{-} = 1/2$. This very simple result cries out for an explanation, 
and here is the simplest one we can furnish. In the case of efficient detection, the  
system state is {\em always} either $\vec{r}_+$ or $\vec{r}_-$, and alternates between them every time a jump occurs. 
Thus, after every jump it finds itself in either of them with the equal probability of $\wp_\pm=1/2$.
We can model the case of imperfect detection, where both decoherence channels have the same efficiency $\eta$ (as we have assumed) as randomly deleting a portion $1-\eta$ of jumps from the full record for perfect efficiency. Since the remaining jumps are an unbiassed sample of the original set of jumps, on average the system state will be equally often in the two states (since the pure post-jump states in the two situations must agree). 

The ensemble average we require can be obtained by calculating the below integral
\beq \label{avbeta2}
{\rm E}^z\bigg[\sqrt{1-\an{\s z}^2} \bigg] = \frac{\sum_{\ell=\pm} \frac{1}{2} \int_0^\infty  p^\ell(t) \sqrt{1-z_\ell^2(t)}\,dt}{\sum_{\ell=\pm} \frac{1}{2} \int_0^\infty  p^\ell(t) \,dt}.
\eeq 
This is directly comparable to Eq.~(A.2). There the time-average was done by numerically simulating a typical trajectory of jumps. Here can calculate exactly the time-average by using the distribution over the initial state $\ell$ (immediately following a jump) and the time  $t$ until the next jump.

 \subsection*{ Simulating Averages for \blk Diffusive Unravellings}
The case we are interested in, where $\Upsilon = \eta \, {\rm diag}(e^{-2i\varphi}, e^{2i\varphi}) $, 
corresponds to to homodyne detection of both channels, with phase $\varphi$. As noted in the main text, 
the ensemble averages are independent of $\varphi$ so without loss of generality we can take $\varphi=0$. 
Then the \blk conditional state of the system evolves according to the following stochastic differential equation
  {\setlength\arraycolsep{2pt}
\begin{eqnarray}   \label{sde}
d\varrho & = & \big( {\cal D}[{\hat{c}}_-]  + {\cal D}[{\hat{c}}_+] \big) \varrho \, dt + \sqrt{\eta} \big( {\cal H} [ {\hat{c}}_- dW_-]  \nonumber \\[4pt]
& & + {\cal H} [ {\hat{c}}_+ dW_+] \big) \varrho \, dt.
\end{eqnarray}}
where $dW_\pm$ are independent real Wiener processes. 
The state of qubit is confined to $y=0$ plane such that at any instant of time it can be identified by a point $\big(x(t), 0, z(t)\big)$ in the Bloch sphere. The evolution of this point is governed by the coupled stochastic differential equations 
  {\setlength\arraycolsep{2pt}
\begin{eqnarray}  {\label {d1}}
dx & = & -(\gamma_{\Sigma}/2) \blk \, x dt + \sqrt{\eta\gamma_-} (1+z-x^2) dW_- \nonumber \\[4pt] 
& & + \sqrt{\eta\gamma_+} (1-z-x^2) dW_+, 
\end{eqnarray}}
  {\setlength\arraycolsep{2pt}
\begin{eqnarray}  {\label {d2}}
dz & = & (-\gamma_{\Sigma} \, z + \gamma_{\Delta})  dt - \sqrt{\eta\gamma_-} x (1+z) dW_- \nonumber \\[4pt] 
& &+ \sqrt{\eta\gamma_+} x (1-z) dW_+.
\end{eqnarray}}
We simulate these using the Milstein method \cite{KloPla92}. \
Once the transients have decayed away (after several $\gamma_\Sigma^{-1}$) 
we record data for both $x$ and $z$, 
to calculate ${\rm E}^\varphi\big[|\langle {\hat{\sigma}}_{\varphi} \rangle|\big] = {\rm E}[|x|]$ and 
${\rm E}^\varphi\sq{\sqrt{1-\langle {\hat{\sigma}}_{z}| \rangle^2}}={\rm E}[\sqrt{1-z^2}]$
as time averages. 

 \subsection*{ Limit of small $R$}
 When $R\equiv\gamma_+/\gamma_- \ll 1$, the conditioned system state under quantum diffusion 
 is almost always near the ground state, and $x = O(\sqrt{R}), 1+z = O({R})$. Then \blk \erfa{d1}{d2}  
 become, to leading order, 
   {\setlength\arraycolsep{2pt}
\begin{eqnarray}  {\label {s1}}
dx & = & -1/2 \, x\, dt + 2 \sqrt{\eta R} \,dW_+ , \\[4pt] {\label {s2}}
dz & = & [ 2R - z - 1]  dt  + 2 \sqrt{\eta R} \, x\, dW_+.
\end{eqnarray}
Under this approximation it is easy to find the first two moments of $x$ and $z$ for the system in steady state:
    {\setlength\arraycolsep{2pt}
\begin{eqnarray}  
{\text E}[x]  & = & 0,  \hspace{40pt} {\text E}[z]  =  2R-1, \\[4pt] 
\text{Var}[x] & = & 4\eta R, \hspace{16pt} \text{Var}[z] = 8 \eta^2 R^2.
\end{eqnarray}

From Eq.~(\ref{s1}) one has a Gaussian distribution for $x$,  which enables us to calculate the first term of the steering parameter,
${\rm E}[|x|]$.  For $z$, however, the above moments show that a Gaussian 
cannot be a good approximation for $z$ (because it is bounded below by $-1$). 
 However, we can consider a Taylor series expansion of $\sqrt{1-z^2}$ about 
 ${\text E}[z]$. This gives 
\blk  the analytic expression of steering parameter for small $R$ as
{\setlength\arraycolsep{2pt}
\begin{equation}   {\label {varz}}
S_{R \ll 1}  \cong   \sqrt{\frac{8 \eta R}{\pi}} - f(n) \Big[ \frac{4 - \eta^2}{2} + h(\eta) \Big] \sqrt{R} \nonumber \\[4pt]
\end{equation} 
where $h(\eta)$ comes from higher-order (beyond second-order) moments of $z$, which are not negligible (they do not scale with $R$). Thus, whatever the form of $h(\eta)$, this does not change the scaling with $R$: 
\beq 
S_{R \ll 1}  \cong g(\eta) \sqrt{R},  
\eeq
 {\rm where} $g(\eta) =  \sqrt{{8 \eta}/{\pi}} - f(n) \Big[({4 - \eta^2})/{2} + h(\eta) \Big].$ 
 Ignoring $h(\eta)$ and using $f(\infty) = 2/\pi$ \cite{JonWis11a}, we predict a critical efficiency, 
 where $g(\eta_c)=0$, of $\eta_c \approx 0.545$. From the stochastic simulations with $R=0.01$, we found 
 (see main text) $\eta_c \approx 0.59$, showing that $h(\eta)$ is non-negligible, as expected, but not very important.


\begin{thebibliography}{99}

   \bibitem{Car93}
H. J.~Carmichael,  \textit{An Open Systems Approach to Quantum Optics} {Springer, Berlin (1993)}.

  \bibitem{DalCasMol92}
J.~Dalibard, Y.~Castin, and K.~M\o lmer \href{http://prl.aps.org/abstract/PRL/v68/i5/p580_1}{Phys. Rev. Lett. {\bf 68}, 580 (1992)}.


\bibitem{GarParZol92}
C. W.~Gardiner, A. S.~Parkins, and P.~Zoller \href{http://pra.aps.org/abstract/PRA/v46/i7/p4363_1}{Phys. Rev. A {\bf 46} 4363 (1992)}.

\bibitem{Bar93}
A.~Barchielli \href{http://link.springer.com/article/10.1007/BF00672994}{Int. J. Theor. Phys. {\bf 32} 2221 (1993)}.

\bibitem{WisMil93c} 
H. M. Wiseman and G. J. Milburn, \href{http://pra.aps.org/abstract/PRA/v47/i3/p1652_1}{Phys. Rev. A {\bf 47}, 1652  (1993)}.

  \bibitem{Boh13}
N.~Bohr \href{http://www.tandfonline.com/doi/abs/10.1080/14786441308634955#.UiaIBhZZuZY} {Philos. Mag. {\bf 26}, 1 (1913)}.

  \bibitem{Ein17}
A.~Einstein {Phys. Z. {\bf 18}, 121 (1917)}.

   \bibitem{GisPer92}
N.~Gisin and I. C.~Percival \href{http://iopscience.iop.org/0305-4470/25/21/023}{J. Phys. A {\bf 25}, 5677 (1992)}.

  \bibitem{WisGam12}
H. M.~Wiseman and J. M.~Gambetta, \href{http://prl.aps.org/abstract/PRL/v108/i22/e220402}{Phys. Rev. Lett. {\bf 108}, 220402 (2012)}.

   \bibitem{EPR35}
A.~Einstein, B.~Podolsky, and N.~Rosen, \href{http://prola.aps.org/abstract/PR/v47/i10/p777_1}{Phys. Rev. {\bf 47}, 777 (1935)}.

  \bibitem{WisJonDoh07}
H. M.~Wiseman, S. J.~Jones, and A. C.~Doherty, \href{http://prl.aps.org/abstract/PRL/v98/i14/e140402}{Phys. Rev. Lett. {\bf 98}, 140402 (2007)}.

\bibitem{CavJonWisRei09}
E. G.~Cavalcanti, S. J.~Jones, H. M.~Wiseman, and M. D.~Reid \href{http://pra.aps.org/abstract/PRA/v80/i3/e032112}{Phys. Rev. A {\bf 80} 032112 (2009)}.

\bibitem{SJWP10}
D. J.~Saunders, S. J.~Jones, H. M.~Wiseman, and G. J.~Pryde \href{http://www.nature.com/nphys/journal/v6/n11/full/nphys1766.html}{Nature Phys. {\bf 6} 845 (2010)}.

   \bibitem{Schro35}
E.~{\sch}, \href{http://journals.cambridge.org/action/displayAbstract?fromPage=online&aid=1737068}{Proc. Cambridge Philos. Soc. {\bf 31}, 553 (1935)}.

  \bibitem{WisGam08}
H. M.~Wiseman and J. M.~Gambetta, \href{http://prl.aps.org/abstract/PRL/v101/i14/e140401}{Phys. Rev. Lett. {\bf 101}, 140401 (2008)}.

  \bibitem{WisMil10}
H. M.~Wiseman and G. J.~Milburn, \textit{Quantum Measurement and Control}, (Cambridge Univ.\ Press, Cambridge, 2010).

\bibitem{Ber05}
D. S.~Bernstein \textit{Matrix Mathematics}, (Princeton Univ.\ Press, Princeton, New Jersey, 2005).

  \bibitem{KarWis11}
R. I.~Karasik and H. M.~Wiseman, \href{http://prl.aps.org/abstract/PRL/v106/i2/e020406}{Phys. Rev. Lett. {\bf 106}, 020406 (2011)}.

  \bibitem{WisVac01}
H. M.~Wiseman and J. A.~Vaccaro, \href{http://prl.aps.org/abstract/PRL/v87/i24/e240402}{Phys. Rev. Lett. {\bf 87}, 240402 (2001)}.

  \bibitem{SanCar12}
M. F.~Santos, M.~Terra Cunha, R.~Chaves and A. R. R.~Carvalho, \href{http://prl.aps.org/abstract/PRL/v108/i17/e170501}{Phys. Rev. Lett. {\bf 108}, 170501 (2012)}.

  \bibitem{JonWis11}
S. J.~Jones and H. M.~Wiseman, \href{http://pra.aps.org/abstract/PRA/v84/i1/e012110}{Phys. Rev. A {\bf 84}, 012110 (2011)}.

\bibitem{Mur13}
K. W.~Murch, S. J.~Weber, C.~Macklin, and I.~Siddiqi, \href{http://www.nature.com/nature/journal/v502/n7470/full/nature12539.html?WT.ec_id=NATURE-20131010}{Nature. {\bf 502}, 211 (2013)}
  
\end{thebibliography}

\begin{thebibliography}{99}


\bibitem[WisMil10]{WisMil10a}
H.~M. Wiseman and G.~J. Milburn, \textit{Quantum Measurement and Control}, (Cambridge Univ.\ Press, Cambridge, 2010).

  \bibitem[JonWis11]{JonWis11a}
S.J.~Jones and H.M.~Wiseman, {Phys. Rev. A  \textbf{84}, 012110 (2011)}.

\bibitem[KloPla92]{KloPla92}
P.E.~Kloeden and E.~Platen, \textit{Numerical Solution of Stochastic Differential Equations}, (Springer-Verlag Berlin Heidelberg, 1992).

\end{thebibliography}
\end{document}